\documentclass[11pt, a4paper]{article}
\usepackage{amssymb}
\usepackage{graphicx}

\begin{document}

\title{Exact solution of the Susceptible-Infectious-Recovered 
(SIR) epidemic model}
\author{Norio Yoshida\thanks{Department of Mathematics, University of Toyama, Toyama, 930-8555 Japan (E-mail: norio.yoshidajp@gmail.com)}}
\date{}

\maketitle
\begin{abstract}
Exact solution of the Susceptible-Infectious-Recovered 
(SIR) epidemic model is derived, and various properties of solution 
are obtained directly from the exact solution. It is shown that 
there exists an exact solution of an initial value problem for 
SIR differential system and that the parametric form of the exact solution 
satisfies some linear differential system. 
\end{abstract}
\vspace{2ex}
\\


\newtheorem{Df}{Definition}
\newtheorem{thm}{Theorem}
\newtheorem{lem}{Lemma}
\newtheorem{cor}{Corollary}
\newtheorem{ex}{Example}
\newtheorem{rem}{Remark}
\newtheorem{prop}{Proposition}
\newcommand{\qed}{\hfill$\Box$}
\font\bi=cmbxti12 
\newcommand{\Proof}{\rm\vspace{1ex}\noindent{\bi Proof.}}

\vspace{2ex}

There is much current requirement for mathematical approach to 
epidemic models.   It goes without saying that a vast literature and 
research papers, dealing with epidemic models has been published so far 
(cf. \cite{bdw08, c93, f80, mr08}).  However, very little is known about 
exact solutions of the epidemic models (cf. \cite{bst19, hlm14, sks10, y22}). 
Motivated by the paper \cite{hlm14} by Harko, Lobo and Mak, 
we establish exact solutions of the Susceptible-Infectious-Recovered (SIR) epidemic differential system.  The purpose of this paper is to obtain an 
exact solution of SIR differential system, and to investigate 
various properties of solution via the exact solution. 
We note that the content of this paper is, for the most part, a special 
case of that of \cite{y22} which deals with SIRD epidemic model. 
However, there will be some meaning in stating for the SIR epidemic model 
which is a very important model in epidemiology. 

In \cite{hlm14} the second order differential equations of the form 
$$
   z'' = x_{0}\beta z'e^{-(\beta/\gamma)z} - \gamma z'
$$
play an important role in investigating exact solutions of 
SIR epidemic model.  
However, in this paper we discuss about the 
exact solution based on some first order nonlinear differential equation. 
Furthermore we show that the parametric form of the exact solution is a 
solution of an initial value problem for 
some linear differential system.

We are concerned with the SIR differential system 
\begin{eqnarray}
   \frac{dS(t)}{dt} & = & -\beta S(t)I(t),  \label{ny1} \\
   \frac{dI(t)}{dt} & = & \beta S(t)I(t) - \gamma I(t), \label{ny2} \\
   \frac{dR(t)}{dt} & = & \gamma I(t) \label{ny3}
\end{eqnarray}
for $t > 0$, where $\beta$ and $\gamma$ are positive constants. 
The initial condition to be considered is the following: 
\begin{equation}
   S(0) = \tilde{S},\ I(0) = \tilde{I},\ R(0) = \tilde{R}, 
   \label{ny4}
\end{equation}
where $\tilde{S} + \tilde{I} + \tilde{R} = N\ (\mbox{positive\ constant})$. 
Since 
$$
   \frac{d}{dt}\left(S(t)+I(t)+R(t)\right) 
   = \frac{dS(t)}{dt}+\frac{dI(t)}{dt}+\frac{dR(t)}{dt} = 0 
$$
by (\ref{ny1})--(\ref{ny3}), it follows that 
$$
   S(t) + I(t) + R(t) = k \ (t \geq 0) 
$$
for some constant $k$.  In view of the fact that 
$$
   k = S(0) + I(0) + R(0) = \tilde{S}+\tilde{I}+\tilde{R}=N,
$$
we conclude that 
$$
   S(t) + I(t) + R(t) = N \ (t \geq 0). 
$$
It is assumed throughout this paper that:
\begin{itemize}
   \item[(A$_{1}$)] $\displaystyle \tilde{S} > \frac{\gamma}{\beta}$; 
   \item[(A$_{2}$)] $\tilde{I} > 0$; 
   \item[(A$_{3}$)] $\tilde{R} \geq 0$ satisfies 
\begin{equation}
   N > \tilde{S}e^{(\beta/\gamma)\tilde{R}} + \tilde{R}. \label{ny5}
\end{equation}
\end{itemize}

\begin{lem} \label{ny:lem1} If $S(t) > 0$ for $t > 0$, then we obtain 
\begin{equation}
   R'(t)
   = \gamma\left(N - \tilde{S}e^{(\beta/\gamma)\tilde{R}}e^{-(\beta/\gamma)R(t)}     - R(t)\right)\ (t > 0). \label{ny6}
\end{equation}
\end{lem}

{\Proof}
It follows from (\ref{ny1}) and (\ref{ny3}) that 
$$
   R'(t) = \gamma I(t) = \gamma\left(\frac{S'(t)}{-\beta S(t)}\right) 
         = - \frac{\gamma}{\beta}\bigl(\log S(t)\bigr)', 
$$
and integrating the above on $[0,t]$ yields 
$$
   R(t) - \tilde{R} 
   = - \frac{\gamma}{\beta}\bigl( \log S(t) - \log \tilde{S}\bigr). 
$$
Therefore we obtain 
$$
   \log S(t) = - \frac{\beta}{\gamma}\bigl(R(t) - \tilde{R}\bigr) 
               + \log \tilde{S} 
$$
and hence
\begin{equation}
   S(t) = \exp \left(\log \tilde{S} - \frac{\beta}{\gamma}R(t) 
                     + \frac{\beta}{\gamma} \tilde{R} \right) 
        = \tilde{S}e^{(\beta/\gamma)\tilde{R}} e^{-(\beta/\gamma)R(t)}. 
          \label{ny7} 
\end{equation}
Since $I(t) = N - S(t) - R(t)$, we observe, using (\ref{ny7}), that 
\begin{eqnarray*}
   R'(t) 
   & = & \gamma I(t) \\
   & = & \gamma (N - S(t) - R(t)) \\
   & = & \gamma \left(N - \tilde{S}e^{(\beta/\gamma)\tilde{R}}
         e^{-(\beta/\gamma)R(t)} - R(t) \right) 
\end{eqnarray*}
which is the desired identity (\ref{ny6}). 
\qed

By a {\it solution} of the SIR differential system (\ref{ny1})--(\ref{ny3}) 
we mean a vector-valued function $(S(t),I(t),R(t))$ of class $C^{1}(0,\infty) \cap C[0,\infty)$ which satisfies (\ref{ny1})--(\ref{ny3}). 
Associated with every continuous function $f(t)$ on $[0,\infty)$, we define
$$
   f(\infty) := \lim_{t \to \infty} f(t).
$$

\begin{lem} \label{ny:lem2} 
Let $(S(t),I(t),R(t))$ be a solution of the 
SIR differential system {\rm (\ref{ny1})--(\ref{ny3})} 
such that $S(t) > 0$ and $I(t) > 0$ for $t > 0$. 
Then there exist the limits $S(\infty),\ I(\infty)$ and $R(\infty)$. 
\end{lem}

{\Proof}
From {\rm (\ref{ny1})} we see that $S'(t) < 0$ for $t > 0$ 
in view of the fact that $S(t)>0$ and $I(t)>0$, and hence 
$S(t)$ is decreasing on $[0,\infty)$. Since $S(t) > 0$, 
it follows that $S(t)$ is bounded from below. 
Therefore there exists $S(\infty)$. 
Since $R'(t) = \gamma I(t) > 0$ and $R(t) \leq N$, we find that $R(t)$ is 
increasing and bounded from above. Hence there exists $R(\infty)$. 
Since $I(t) = N - S(t) - R(t)$ and there exist $S(\infty)$ and $R(\infty)$, 
we see that there exists $I(\infty)$. 
\qed

The following theorem is due to Harko, Lobo and Mak \cite{hlm14}. 

\begin{thm} \label{ny:thm1}
Let $(S(t),I(t),R(t))$ be a solution of the initial value problem 
{\rm (\ref{ny1})--(\ref{ny4})} such that $S(t) > 0$ and 
$I(t) > 0$ for $t > 0$. 
Then $(S(t),I(t),R(t))$ can be represented in the following parametric form:
\begin{eqnarray}
   S(t) & = & S(\varphi(u))         
          = \tilde{S}e^{(\beta/\gamma)\tilde{R}}u, \label{ny8} \\
   I(t) & = & I(\varphi(u)) 
          = \frac{\gamma}{\beta}\,\log u 
      - \tilde{S}e^{(\beta/\gamma)\tilde{R}}u + N, \label{ny9} \\ 
   R(t) & = & R(\varphi(u)) 
          =  - \frac{\gamma}{\beta}\,\log u \label{ny10}
\end{eqnarray}
for $e^{-(\beta/\gamma)R(\infty)} < u \leq e^{-(\beta/\gamma)\tilde{R}}$, 
where $t = \varphi(u)$ is given in the proof. 
\end{thm}

{\Proof}
We define the function $u(t)$ by
$$
   u(t) := e^{-(\beta/\gamma)R(t)}. 
$$
Then $u = u(t)$ is decreasing on $[0,\infty)$, 
$e^{-(\beta/\gamma)R(\infty)} < u \leq e^{-(\beta/\gamma)\tilde{R}}$ 
and $\lim\limits_{t \to \infty} u(t) = e^{-(\beta/\gamma)R(\infty)}$ 
because $R(t)$ is increasing on $[0,\infty)$ and 
$\tilde{R} \leq R(t) < R(\infty)$. It is obvious that 
$u(t)$ is of class $C^{1}(0,\infty)$ in view of 
$e^{-(\beta/\gamma)R(t)} \in C^{1}(0,\infty)$. 
Hence, there exists the inverse function 
$\varphi(u) \in 
C^{1}(e^{-(\beta/\gamma)R(\infty)},e^{-(\beta/\gamma)\tilde{R}})$ 
of $u = u(t)$ 
such that 
$$
   t = \varphi(u) \quad
   \left(e^{-(\beta/\gamma)R(\infty)} < u \leq 
   e^{-(\beta/\gamma)\tilde{R}}\right), 
$$
$\varphi(u)$ is decreasing in 
$(e^{-(\beta/\gamma)R(\infty)}, e^{-(\beta/\gamma)\tilde{R}}]$, 
$\varphi\bigl(e^{-(\beta/\gamma)\tilde{R}}\bigr) = 0$ and \\
$\lim_{u \to e^{-(\beta/\gamma)R(\infty)}+0} \varphi(u) = \infty$. 
Substituting $t = \varphi(u)$ into (\ref{ny6}), we derive 
\begin{equation}
   R'(\varphi(u))
   = \gamma\left(N - \tilde{S}e^{(\beta/\gamma)\tilde{R}}
     e^{-(\beta/\gamma)R(\varphi(u))} 
     - R(\varphi(u))\right).  \label{ny11}
\end{equation}
Differentiating both sides of 
$\displaystyle u = e^{-(\beta/\gamma)R(\varphi(u))}$ 
with respect to $u$, we obtain 
\begin{eqnarray*}
   1 & = & -\frac{\beta}{\gamma}R'(\varphi(u))\varphi'(u)
           e^{-(\beta/\gamma)R(\varphi(u))} \\
     & = & -\frac{\beta}{\gamma}R'(\varphi(u))\varphi'(u)u, 
\end{eqnarray*}
and therefore 
\begin{equation}
   R'(\varphi(u)) = -\frac{\gamma}{\beta} \frac{1}{\varphi'(u)u}. 
   \label{ny12}
\end{equation}
It is clear that 
\begin{equation}
   R(\varphi(u)) = - \frac{\gamma}{\beta} \log u  \label{ny13}
\end{equation}
in light of $u = e^{-(\beta/\gamma)R(\varphi(u))}$. 
Combining (\ref{ny11})--(\ref{ny13}) yields 
$$
   -\frac{\gamma}{\beta}\frac{1}{\varphi'(u)u} 
   = \gamma N - \gamma \tilde{S}e^{(\beta/\gamma)\tilde{R}}u 
     + \frac{\gamma^{2}}{\beta}\log u 
$$
and hence
\begin{eqnarray}
   \varphi'(u) & = & -\frac{\gamma}{\beta}\,
   \frac{1}{u\left(\gamma N - \gamma \tilde{S}e^{(\beta/\gamma)\tilde{R}}u
   + \frac{\gamma^{2}}{\beta}\log u\right) }  \nonumber \\
   & = & \frac{1}{u\left(-\beta N + \beta \tilde{S}e^{(\beta/\gamma)\tilde{R}}u
         - \gamma \log u \right)}.   \label{ny14}
\end{eqnarray}
Integrating (\ref{ny14}) over 
$[u, e^{-(\beta/\gamma)\tilde{R}}]$, we arrive at
$$
   \int_{u}^{e^{-(\beta/\gamma)\tilde{R}}} \varphi'(\xi)d\xi 
   = \int_{u}^{e^{-(\beta/\gamma)\tilde{R}}} 
     \frac{d\xi}{\xi\left(-\beta N 
     + \beta \tilde{S}e^{(\beta/\gamma)\tilde{R}}\xi
         - \gamma \log \xi \right)}. 
$$
Since the left hand side of the above identity is equal to 
$$
    \varphi(e^{-(\beta/\gamma)\tilde{R}}) - \varphi(u)
   = - \varphi(u), 
$$
we are led to 
\begin{eqnarray*}
   t = \varphi(u) & = & \int_{u}^{e^{-(\beta/\gamma)\tilde{R}}} 
     \frac{d\xi}{\xi\left(\beta N 
     - \beta \tilde{S}e^{(\beta/\gamma)\tilde{R}}\xi
         + \gamma \log \xi \right)} \\
   & = & \int_{u}^{e^{-(\beta/\gamma)\tilde{R}}}
         \frac{d\xi}{\xi\psi(\xi)}, 
\end{eqnarray*}
where
\begin{equation}
   \psi(\xi) = \beta N 
     - \beta \tilde{S}e^{(\beta/\gamma)\tilde{R}}\xi
         + \gamma \log \xi. \label{ny15}
\end{equation}
From (\ref{ny7}) and (\ref{ny13}) it follows that
\begin{eqnarray*}
   S(t) & = & S(\varphi(u)) 
          = \tilde{S}e^{(\beta/\gamma)\tilde{R}}
            e^{-(\beta/\gamma)R(\varphi(u))} 
          =  \tilde{S}e^{(\beta/\gamma)\tilde{R}}u, \\
   R(t) & = & R(\varphi(u)) 
          = - \frac{\gamma}{\beta} \log u, \\
   I(t) & = & I(\varphi(u)) 
          = N - S(\varphi(u)) - R(\varphi(u)) \\
        &  = & N - \tilde{S}e^{(\beta/\gamma)\tilde{R}}u 
            + \frac{\gamma}{\beta} \log u, 
\end{eqnarray*}
which is the desired solution (\ref{ny8})--(\ref{ny10}). 
Since $\lim_{u \to e^{-(\beta/\gamma)R(\infty)}+0} \varphi(u) = \infty$, 
it is necessary that 
\begin{eqnarray*}
   \lim_{\xi \to e^{-(\beta/\gamma)R(\infty)}+0}\psi(\xi)
   & = & \lim_{\xi \to e^{-(\beta/\gamma)R(\infty)}+0} 
         \bigl(\beta N 
         - \beta \tilde{S}e^{(\beta/\gamma)\tilde{R}}\xi 
         + \gamma \log \xi \bigr) \\
   & = & \lim_{x \to R(\infty)-0} 
         \beta \bigl( N 
         - \tilde{S}e^{(\beta/\gamma)\tilde{R}}
         e^{-(\beta/\gamma)x} - x \bigr) \\
   & = & \beta\left(N - \tilde{S}e^{(\beta/\gamma)\tilde{R}}
         e^{-(\beta/\gamma)R(\infty)} - R(\infty) \right) \\
   & = & 0, 
\end{eqnarray*}
which implies 
$$
   N - \tilde{S}e^{(\beta/\gamma)\tilde{R}}
     e^{-(\beta/\gamma)R(\infty)} - R(\infty) = 0. 
$$
We see that $\psi'(\xi) = 0$ for 
$\xi = \tilde{\xi} = 
\bigl(\gamma/(\beta\tilde{S})\bigr)e^{-(\beta/\gamma)\tilde{R}}$, 
and that $e^{-(\beta/\gamma)R(\infty)} < \tilde{\xi} 
< e^{-(\beta/\gamma)\tilde{R}}$ if 
$\gamma/\beta < \tilde{S} < 
(\gamma/\beta)e^{(\beta/\gamma)(R(\infty)-\tilde{R})}$. 
Since $\psi'(\xi) > 0$ for $e^{-(\beta/\gamma)R(\infty)} < \xi < \tilde{\xi}$ 
and $\psi'(\xi) < 0$ for $\tilde{\xi} < \xi < e^{-(\beta/\gamma)\tilde{R}}$, 
it follows that $\psi(\xi)$ is increasing in 
$(e^{-(\beta/\gamma)R(\infty)}, \tilde{\xi})$ and is decreasing in 
$(\tilde{\xi}, e^{-(\beta/\gamma)\tilde{R}})$. 
In view of the fact that 
$\psi\bigl(e^{-(\beta/\gamma)\tilde{R}}\bigr) 
= \beta\left(N - \tilde{S} - \tilde{R}\right) 
= \beta \tilde{I} > 0$ and 
$\lim_{\xi \to e^{-(\beta/\gamma)R(\infty)}+0}\psi(\xi) = 0$, 
we conclude that 
$$
   \psi(\xi) > 0 \quad \mbox{in}\  (e^{-(\beta/\gamma)R(\infty)}, 
   e^{-(\beta/\gamma)\tilde{R}}] 
$$
under the condition 
$\gamma/\beta < \tilde{S} < 
   (\gamma/\beta)e^{(\beta/\gamma)(R(\infty)-\tilde{R})}$. 
Furthermore we obtain 
\begin{eqnarray*}
   \lim_{t \to \infty} \frac{\beta}{\gamma}R'(t) 
   & = & \lim_{t\to\infty} 
         \beta\left(N - \tilde{S}e^{(\beta/\gamma)\tilde{R}}
         e^{-(\beta/\gamma)R(t)} - R(t)\right) \\
   & = & \beta\left(N - \tilde{S}e^{(\beta/\gamma)\tilde{R}}
         e^{-(\beta/\gamma)R(\infty)} - R(\infty) \right) \\
   & = & 0, 
\end{eqnarray*}
which implies $I(\infty) = 0$ by (\ref{ny3}). 
\qed

\begin{cor} \label{ny:cor1} 
Let $(S(t),I(t),R(t))$ be a solution of the initial value problem 
{\rm (\ref{ny1})--(\ref{ny4})} such that $S(t) > 0$ and 
$I(t) > 0$ for $t > 0$. 
Then we have 
\begin{eqnarray*}
   S(t) & = & \tilde{S}e^{(\beta/\gamma)\tilde{R}}e^{-(\beta/\gamma)R(t)} \\
   I(t) & = & - R(t) - \tilde{S}e^{(\beta/\gamma)\tilde{R}}
              e^{-(\beta/\gamma)R(t)} + N. 
\end{eqnarray*}
\end{cor}

{\Proof} 
Since $u = \varphi^{-1}(t) = e^{-(\beta/\gamma)R(t)}$ in the proof of 
Theorem \ref{ny:cor1}, we get 
\begin{eqnarray*}
   S(t) & = & \tilde{S}e^{(\beta/\gamma)\tilde{R}}\varphi^{-1}(t) 
              = \tilde{S}e^{(\beta/\gamma)\tilde{R}}e^{-(\beta/\gamma)R(t)}, \\
   I(t) & = & \frac{\gamma}{\beta}\,\log \varphi^{-1}(t) 
              - \tilde{S}e^{(\beta/\gamma)\tilde{R}}\varphi^{-1}(t) + N \\
        & = & - R(t) - \tilde{S}e^{(\beta/\gamma)\tilde{R}}
              e^{-(\beta/\gamma)R(t)} + N. 
\end{eqnarray*}
\qed

\begin{lem} \label{ny:lem3} Under 
the hypothesis {\rm (A$_{3}$)}, 
the transcendental equation 
\begin{equation}
   x = N - \tilde{S}e^{(\beta/\gamma)\tilde{R}}
                 e^{-(\beta/\gamma)x} \label{ny16}
\end{equation}
has a unique solution $x = \alpha$ such that
$$
   \tilde{R} < \alpha < N 
$$
{\rm ({\it cf}. {\sc Figure} 1)}. 
\end{lem}

%Figure 1
\begin{figure}[h]
  \begin{center}
    \includegraphics[height=5cm]{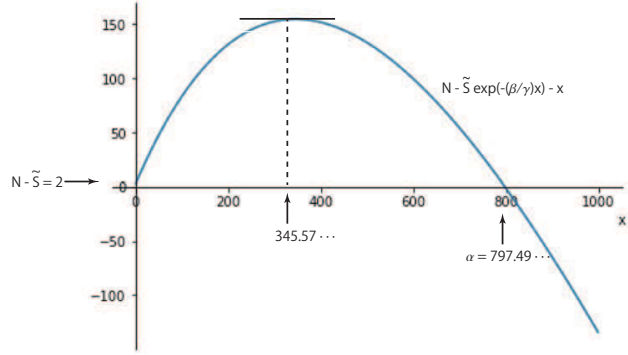}
    \caption{Variation of $N-\tilde{S}e^{-(\beta/\gamma)x}-x$ 
    for $N=1000, \tilde{S}=998, \tilde{R}=0, \beta=0.2/1000$ and $\gamma=0.1$. 
    In this case we find that $N-\tilde{S}=2$ and 
    $0 < \alpha = 797.49\cdots < 1000$. }
  \end{center}
\end{figure}

{\Proof}
It follows from (\ref{ny5}) that 
$$
   N - \tilde{S}e^{(\beta/\gamma)\tilde{R}} > 0. 
$$
Define the sequence $\{a_{n}\}_{n=1}^{\infty}$ by
\begin{eqnarray}
   a_{1}   & = & \tilde{a}\ 
                 \left(0 < \tilde{a} \leq N 
                 - \tilde{S}e^{(\beta/\gamma)\tilde{R}}\right), \nonumber\\
   a_{n+1} & = & N - \tilde{S}e^{(\beta/\gamma)\tilde{R}} 
                 e^{-(\beta/\gamma)a_{n}}\ (n=1,2,...). \label{ny17}
\end{eqnarray}
It is easy to check that 
$$
   a_{1} = \tilde{a} \leq N - \tilde{S}e^{(\beta/\gamma)\tilde{R}}
   \leq N - \tilde{S}e^{(\beta/\gamma)\tilde{R}}e^{-\frac{\beta}{\gamma}a_{1}}
   = a_{2}. 
$$
If $a_{n+1} \geq a_{n}$, then 
\begin{eqnarray*}
   a_{n+2} - a_{n+1} 
   & = & N - \tilde{S}e^{(\beta/\gamma)\tilde{R}}
         e^{-(\beta/\gamma)a_{n+1}} 
         - \left(N - \tilde{S}e^{(\beta/\gamma)\tilde{R}}
         e^{-(\beta/\gamma)a_{n}}\right) \\
   & = & \tilde{S}e^{(\beta/\gamma)\tilde{R}}
         \left(e^{-(\beta/\gamma)a_{n}}-e^{-(\beta/\gamma)a_{n+1}}
         \right) \\
   & \geq & 0. 
\end{eqnarray*}
Therefore we have $a_{n+2} \geq a_{n+1}$, and hence the sequence $\{a_{n}\}$ 
is nondecreasing by the mathematical induction. 
We easily see that the sequence $\{a_{n}\}$ 
is bounded because 
$$
   \vert a_{n+1} \vert \leq N + \tilde{S}e^{(\beta/\gamma)\tilde{R}}
                          e^{-(\beta/\gamma)a_{n}} 
   \leq N + \tilde{S}e^{(\beta/\gamma)\tilde{R}}. 
$$
Consequently there exists $\lim_{n \to \infty} a_{n} = \alpha$. 
Taking the limit as $n \to \infty$ in (\ref{ny17}), we obtain 
$$
   \alpha = N - \tilde{S}e^{(\beta/\gamma)\tilde{R}}
                 e^{-(\beta/\gamma)\alpha}. 
$$
The uniqueness of $\alpha$ follows the fact that 
the straight line $y = N-x$ and the exponential curve 
$y = \tilde{S}e^{(\beta/\gamma)\tilde{R}}e^{-(\beta/\gamma)x}$ 
has only one intersecting point in $0 < x < N$ because of 
the inequality $N > \tilde{S}e^{(\beta/\gamma)\tilde{R}}$, 
and hence the solution $\alpha$ of 
the transcendental equation (\ref{ny16}) is unique. 
The inequality $\tilde{R} < \alpha < N$ follows from 
\begin{eqnarray*}
   \alpha & = & N - \tilde{S}e^{(\beta/\gamma)\tilde{R}}
                e^{-(\beta/\gamma)\alpha} < N, \\
   \alpha & = & N - \tilde{S}e^{(\beta/\gamma)\tilde{R}}
                e^{-(\beta/\gamma)\alpha}
                > N - \tilde{S}e^{(\beta/\gamma)\tilde{R}} > \tilde{R}.
\end{eqnarray*}
\qed

We assume that the following hypothesis
\begin{itemize}
   \item[(A$_{4}$)] $\displaystyle \tilde{S} < 
   \frac{\gamma}{\beta}e^{(\beta/\gamma)(\alpha-\tilde{R})}$ 
\end{itemize}
holds in the rest of this paper. 
We note that (A$_{4}$) is equivalent to the following 
\begin{itemize}
   \item[(A$_{4}'$)] $\displaystyle \frac{\gamma}{\beta} > N - \alpha$ 
\end{itemize}
in light of $\tilde{S}e^{(\beta/\gamma)\tilde{R}}e^{-(\beta/\gamma)\alpha} 
= N - \alpha$. 

\begin{thm} \label{ny:thm2}
The initial value problem {\rm (\ref{ny1})--(\ref{ny4})} has 
the solution
\begin{eqnarray}
   S(t) & = & \tilde{S}e^{(\beta/\gamma)\tilde{R}}\varphi^{-1}(t), 
              \label{ny18} \\
   I(t) & = & \frac{\gamma}{\beta}\,\log \varphi^{-1}(t)
              - \tilde{S}e^{(\beta/\gamma)\tilde{R}}\varphi^{-1}(t)
              + N,   \label{ny19} \\
   R(t) & = & - \frac{\gamma}{\beta}\,\log \varphi^{-1}(t), \label{ny20}
\end{eqnarray}
where $\varphi^{-1}(t)$ denotes the inverse function of 
$\varphi : (e^{-(\beta/\gamma)\alpha}, e^{-(\beta/\gamma)\tilde{R}}]  
\rightarrow [0,\infty)$ such that 
$$
     t = \varphi(u)  :=  \int_{u}^{e^{-(\beta/\gamma)\tilde{R}}} 
     \frac{d\xi}{\xi\psi(\xi)}. 
$$
\end{thm}

{\Proof}
First we note that $\varphi(u) \in 
C^{1}(e^{-(\beta/\gamma)\alpha},e^{-(\beta/\gamma)\tilde{R}})$, 
$\varphi(u)$ is decreasing in 
$(e^{-(\beta/\gamma)\alpha}, e^{-(\beta/\gamma)\tilde{R}}]$, 
$\varphi\bigl(e^{-(\beta/\gamma)\tilde{R}}\bigr) = 0$. 
We claim that $\lim_{u \to e^{-(\beta/\gamma)\alpha}+0} \varphi(u) 
= \infty$. 
A simple computation shows that
\begin{eqnarray}
   \lim_{\xi \to e^{-(\beta/\gamma)\alpha}+0}\psi(\xi)
   & = & \lim_{\xi \to e^{-(\beta/\gamma)\alpha}+0} 
         \bigl( \beta N 
         - \beta \tilde{S}e^{(\beta/\gamma)\tilde{R}} \xi 
         + \gamma \log \xi \bigr)  \nonumber \\
   & = & \lim_{x \to \alpha -0} 
         \beta \bigl( N - \tilde{S}e^{(\beta/\gamma)\tilde{R}}
         e^{-(\beta/\gamma)x} - x \bigr)  \nonumber \\
   & = & \beta\bigl(N - \tilde{S}e^{(\beta/\gamma)\tilde{R}}
         e^{-(\beta/\gamma)\alpha} - \alpha \bigr) \nonumber \\
   & = & 0.  \label{ny21}
\end{eqnarray}
Taking into account the hypotheses (A$_{1}$) and (A$_{4}$), 
we observe that 
$\psi(\xi) > 0$ in 
$(e^{-(\beta/\gamma)\alpha}, e^{-(\beta/\gamma)\tilde{R}}]$ 
by the same arguments as in the proof of Theorem \ref{ny:thm1}. 
Since 
$$
   \frac{1}{\xi\psi(\xi)} 
    =  \frac{\beta}{\gamma}\tilde{S}e^{(\beta/\gamma)\tilde{R}}
         \frac{1}{\psi(\xi)} 
         + \frac{1}{\gamma}\,\frac{- \beta \tilde{S}e^{(\beta/\gamma)\tilde{R}}
           + (\gamma/\xi)}{\psi(\xi)},
$$
we obtain 
\begin{eqnarray}
   \varphi(u)
   & = & \int_{u}^{e^{-(\beta/\gamma)\tilde{R}}} 
         \frac{d\xi}{\xi\psi(\xi)}  \nonumber \\
   & = & \frac{\beta}{\gamma}\tilde{S}e^{(\beta/\gamma)\tilde{R}}
         \int_{u}^{e^{-(\beta/\gamma)\tilde{R}}} \frac{d\xi}{\psi(\xi)} 
         + \frac{1}{\gamma}\int_{u}^{e^{-(\beta/\gamma)\tilde{R}}} 
         \frac{\psi'(\xi)}{\psi(\xi)}d\xi  \nonumber \\
   & = & \frac{\beta}{\gamma}\tilde{S}e^{(\beta/\gamma)\tilde{R}}
         \int_{u}^{e^{-(\beta/\gamma)\tilde{R}}} \frac{d\xi}{\psi(\xi)} 
         + \frac{1}{\gamma}\left(
         \log\,\psi\bigl(e^{-(\beta/\gamma)\tilde{R}}\bigr) 
         - \log\,\psi(u) \right)  \nonumber \\
   & = & \frac{\beta}{\gamma}\tilde{S}e^{(\beta/\gamma)\tilde{R}}
         \int_{u}^{e^{-(\beta/\gamma)\tilde{R}}} \frac{d\xi}{\psi(\xi)} 
         + \frac{1}{\gamma}\left(
         \log\,(\beta \tilde{I}) - \log\,\psi(u) \right). \label{ny22}
\end{eqnarray}
Hence it follows from (\ref{ny21}) and (\ref{ny22}) that 
$$
   \lim_{u \to e^{-(\beta/\gamma)\alpha}+0} \varphi(u) 
   = \lim_{u \to e^{-(\beta/\gamma)\alpha}+0} 
     \int_{u}^{e^{-(\beta/\gamma)\tilde{R}}} 
     \frac{d\xi}{\xi\psi(\xi)} = \infty. 
$$
Then we deduce that 
$\varphi^{-1}(t) \in C^{1}(0,\infty)$, $\varphi^{-1}(t)$ is 
decreasing on $[0,\infty)$, and that
\begin{eqnarray*}
   & & \varphi^{-1}(0) = e^{-(\beta/\gamma)\tilde{R}}, \\
   & & \lim_{t \to \infty} \varphi^{-1}(t) = e^{-(\beta/\gamma)\alpha}. 
\end{eqnarray*}
It is easy to check that 
\begin{eqnarray}
   \left(\varphi^{-1}(t)\right)' 
   & = & \frac{1}{\varphi'(u)}\Big\vert_{u=\varphi^{-1}(t)} \nonumber \\
   & = & u\left(-\beta N + \beta \tilde{S}e^{(\beta/\gamma)\tilde{R}}u
         - \gamma \log u \right)\Big\vert_{u=\varphi^{-1}(t)} \nonumber \\
   & = & \varphi^{-1}(t)\left(-\beta N + \beta 
         \tilde{S}e^{(\beta/\gamma)\tilde{R}}\varphi^{-1}(t)
         - \gamma \log \varphi^{-1}(t) \right)  \nonumber \\
   & = & - \varphi^{-1}(t)\psi\bigl(\varphi^{-1}(t)\bigr) = 
         - \beta I(t)\varphi^{-1}(t).  \label{ny23}
\end{eqnarray}
Now we show that (\ref{ny18})--(\ref{ny20}) satisfies 
(\ref{ny1})--(\ref{ny4}). 
Using (\ref{ny18})--(\ref{ny20}) and (\ref{ny23}), we arrive at
\begin{eqnarray*}
   S'(t) 
   & = & \tilde{S}e^{(\beta/\gamma)\tilde{R}}
         \left(\varphi^{-1}(t)\right)' 
         =  \tilde{S}e^{(\beta/\gamma)\tilde{R}} 
         \left(- \beta I(t)\varphi^{-1}(t)\right) \\
   & = & - \beta \tilde{S}e^{(\beta/\gamma)\tilde{R}}
         \varphi^{-1}(t)I(t) 
         =  - \beta S(t)I(t), \\
   I'(t) 
   & = & - \tilde{S}e^{(\beta/\gamma)\tilde{R}}
         \left(\varphi^{-1}(t)\right)' 
         + \frac{\gamma}{\beta} 
         \frac{\left(\varphi^{-1}(t)\right)'}{\varphi^{-1}(t)} \\
   & = & - (- \beta S(t)I(t)) + \frac{\gamma}{\beta}(- \beta I(t)) 
         =  \beta S(t)I(t) - \gamma I(t), \\
   R'(t) 
   & = & - \frac{\gamma}{\beta} 
         \frac{\left(\varphi^{-1}(t)\right)'}{\varphi^{-1}(t)} 
         =  - \frac{\gamma}{\beta}(- \beta I(t)) 
         =  \gamma I(t). 
\end{eqnarray*} 
We easily obtain 
\begin{eqnarray*}
   S(0) & = & \tilde{S}e^{(\beta/\gamma)\tilde{R}}\varphi^{-1}(0) 
              = \tilde{S}e^{(\beta/\gamma)\tilde{R}}
                e^{-(\beta/\gamma)\tilde{R}}
              =  \tilde{S},  \\
   R(0) & = & - \frac{\gamma}{\beta}\,\log \varphi^{-1}(0) 
              =  - \frac{\gamma}{\beta}\,\log e^{-(\beta/\gamma)\tilde{R}} 
              =  \tilde{R},  \\
   I(0) & = & \frac{\gamma}{\beta}\,\log \varphi^{-1}(0)
              - \tilde{S}e^{(\beta/\gamma)\tilde{R}}\varphi^{-1}(0) + N 
              =  - \tilde{R} - \tilde{S} + N 
              =  \tilde{I}. 
\end{eqnarray*}
We remark that $\varphi^{-1}(t)$ depends on 
$\tilde{S},\ \tilde{R},\ \tilde{I}$ in light of 
$N = \tilde{S} + \tilde{R} + \tilde{I}$. 
\qed

%Figure 2
\begin{figure}[h]
  \begin{center}
    \includegraphics[height=5cm]{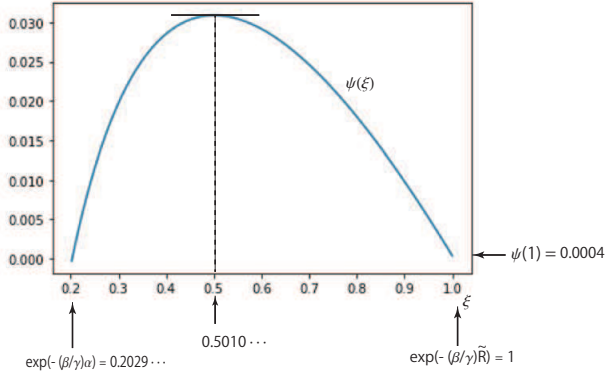}
    \caption{Variation of 
    $\psi(\xi)=\beta N-\beta\tilde{S}e^{(\beta/\gamma)\tilde{R}}\xi
    +\gamma\log \xi$ 
    for $N=1000, \tilde{S}=998, \tilde{R}=0, \beta=0.2/1000$ and $\gamma=0.1$. 
    In this case we see that $e^{-(\beta/\gamma)\alpha} = 0.2029\cdots$, 
    $e^{-(\beta/\gamma)\tilde{R}} = 1$, 
    $\psi(1)=\beta N-\beta\tilde{S} = 0.0004$, $\psi'(\xi) = 0$ for 
    $\xi=0.5010\cdots$, $\gamma/\beta = 500$, 
    $(\gamma/\beta)e^{(\beta/\gamma)\alpha} > 2461$, and 
    $(\gamma/\beta) < \tilde{S} < (\gamma/\beta)e^{(\beta/\gamma)\alpha}$.}    \end{center}
\end{figure}

We can obtain various properties of solution of SIR model via 
the differential system qualitatively, however we derive 
more detailed properties directly from 
the exact solution of the SIR differential system. 

\begin{thm} \label{ny:thm3}
We find that 
$I(\infty) = 0$ and $I(t) > 0$ on $[0,\infty)$, 
and that 
$I(t)$ has the maximum 
$$
   \max_{t \geq 0} I(t) = N - \tilde{R} 
         - \frac{\gamma}{\beta}\left(1 + \log\,\tilde{S} 
         - \log\,\frac{\gamma}{\beta}\right) 
$$
at 
$$
   t = T := \varphi\left(\frac{\gamma}{\beta \tilde{S}
            e^{(\beta/\gamma)\tilde{R}}} \right) 
          = S^{-1}\left(\frac{\gamma}{\beta}\right). 
$$
Furthermore, 
$I(t)$ is increasing in $[0, T)$ and is decreasing in $(T, \infty)$. 
\end{thm}

{\Proof}
It is easily seen that 
\begin{eqnarray*}
   I(\infty) 
   & = & \lim_{t \to \infty} I(t) \\
   & = & \lim_{u \to e^{-(\beta/\gamma)\alpha} + 0} 
         \left(\frac{\gamma}{\beta}\,\log u 
         - \tilde{S}e^{(\beta/\gamma)\tilde{R}}u
         + N \right) \\
   & = & - \alpha - \tilde{S}e^{(\beta/\gamma)\tilde{R}}
         e^{-(\beta/\gamma)\alpha} + N \\
   & = & 0. 
\end{eqnarray*}
Combining (\ref{ny15}) with (\ref{ny19}) yields
\begin{equation}
   I(t) = \frac{1}{\beta} \psi\bigl(\varphi^{-1}(t)\bigr). 
   \label{ny24}
\end{equation}
Since $e^{-(\beta/\gamma)\alpha} < \varphi^{-1}(t) \leq 
e^{-(\beta/\gamma)\tilde{R}}$ for $t \geq 0$ and 
$\psi(\xi) > 0$ for $e^{-(\beta/\gamma)\alpha} < \xi 
\leq e^{-(\beta/\gamma)\tilde{R}}$, 
we find that $I(t) = (1/\beta) \psi\bigl(\varphi^{-1}(t)\bigr) > 0$ 
on $[0,\infty)$. 
Differentiating both sides of (\ref{ny24}) and 
taking account of (\ref{ny23}), we obtain 
\begin{eqnarray}
  I'(t) & = & \frac{1}{\beta} 
              \left( - \beta \tilde{S}e^{(\beta/\gamma)\tilde{R}} 
              + \frac{\gamma}{\varphi^{-1}(t)} \right)
               \bigl(\varphi^{-1}(t)\bigr)'  \nonumber \\
   & = & \frac{1}{\beta}
         \left(- \beta S(t) + \gamma\right) 
         \frac{\bigl(\varphi^{-1}(t)\bigr)'}{\varphi^{-1}(t)} \nonumber \\
   & = & \frac{1}{\beta} 
         \left(- \beta S(t) + \gamma\right)(- \beta I(t))  \nonumber \\
   & = & \bigl(\beta S(t) - \gamma\bigr)I(t). \label{ny25}
\end{eqnarray}
It is easy to see that $I'(t) = 0$ if and only if 
$$
   \varphi^{-1}(t) = \frac{\gamma}{\beta \tilde{S}
            e^{(\beta/\gamma)\tilde{R}}}
$$
or 
$$
   S(t) = \frac{\gamma}{\beta}, 
$$
which yield
$$
   t = T = \varphi\left(\frac{\gamma}{\beta \tilde{S}
            e^{(\beta/\gamma)\tilde{R}}} \right) 
         = S^{-1}\left(\frac{\gamma}{\beta}\right). 
$$
We note that 
$$
   e^{-(\beta/\gamma)\alpha} 
   < \frac{\gamma}{\beta \tilde{S}e^{(\beta/\gamma)\tilde{R}}} 
   = \frac{\gamma}{\beta \tilde{S}}e^{-(\beta/\gamma)\tilde{R}}
   < e^{-(\beta/\gamma)\tilde{R}}
$$
in light of the hypotheses (A$_{1}$) and (A$_{4}$). 
Since $\bigl(\varphi^{-1}(t)\bigr)' < 0$ and 
$\varphi^{-1}(t)$ is decreasing on $[0,\infty)$, we observe that 
$I'(t) > 0\ [\mbox{resp.} < 0]$ if and only if 
$t < T\ [\mbox{resp.} > T]$. 
Therefore, $I(t)$ is increasing in $[0, T)$ and is decreasing 
in $(T, \infty)$.
It is easy to see that the maximum of $I(t)$ on $[0,\infty)$ is given by 
\begin{eqnarray*}
   \frac{1}{\beta} \psi\left(\frac{\gamma}{\beta \tilde{S}
            e^{(\beta/\gamma)\tilde{R}}}\right) 
   & = & N - \tilde{S}e^{(\beta/\gamma)\tilde{R}}
         \frac{\gamma}{\beta \tilde{S}
         e^{(\beta/\gamma)\tilde{R}}}
         + \frac{\gamma}{\beta}
         \log \left(\frac{\gamma}{\beta \tilde{S}
         e^{(\beta/\gamma)\tilde{R}}}\right) \\
   & = & N - \frac{\gamma}{\beta} 
         + \frac{\gamma}{\beta}\left(\log\, \frac{\gamma}{\beta} 
         - \log\,\tilde{S} - \frac{\beta}{\gamma}\tilde{R}\right) \\
   & = & N - \tilde{R} 
         - \frac{\gamma}{\beta}\left(1 + \log\,\tilde{S} 
         - \log\,\frac{\gamma}{\beta}\right).
\end{eqnarray*}
\qed

\begin{cor} \label{ny:cor2} 
The function $I(t)$ has the maximum at 
\begin{eqnarray*}
   T & = & \varphi\left(\frac{\gamma}{\beta \tilde{S}
           e^{(\beta/\gamma)\tilde{R}}} \right)  \\
     & = & \frac{\beta}{\gamma}\tilde{S}e^{(\beta/\gamma)\tilde{R}}
           \int_{\gamma/(\beta \tilde{S}
           e^{(\beta/\gamma)\tilde{R}})}^{e^{-(\beta/\gamma)\tilde{R}}} 
           \frac{d\xi}{\psi(\xi)} 
           + \frac{1}{\gamma}\left(\log\,(\beta \tilde{I}) 
           - \log\,\bigl(\beta F(N,\tilde{S},\tilde{R},\beta,\gamma)\bigr) 
           \right), 
\end{eqnarray*}
and $T$ satisfies the following inequality 
$$
   \tau_{1} \leq T \leq \tau_{2}, 
$$
where
$$
   \tau_{1} 
    =  \frac{(\beta/\gamma)\tilde{S}-1}
         {\beta F(N,\tilde{S},\tilde{R},\beta,\gamma)} 
        + \frac{1}{\gamma}\log\,(\beta \tilde{I}) 
        - \frac{1}{\gamma}
        \log\,\bigl(\beta F(N,\tilde{S},\tilde{R},\beta,\gamma)\bigr) 
$$
and 
$$
   \tau_{2} 
    =  \frac{(\beta/\gamma)\tilde{S}-1}{\beta \tilde{I}} 
        + \frac{1}{\gamma}\log\,(\beta \tilde{I}) - \frac{1}{\gamma}
        \log\,\bigl(\beta F(N,\tilde{S},\tilde{R},\beta,\gamma)\bigr) 
$$
with $F(N,\tilde{S},\tilde{R},\beta,\gamma)$ being
$$
   F(N,\tilde{S},\tilde{R},\beta,\gamma) 
   := N - \tilde{R} 
      - \frac{\gamma}{\beta}\left(1 + \log\,\tilde{S} 
      - \log\,\frac{\gamma}{\beta}\right). 
$$
\end{cor}

{\Proof}
Using (\ref{ny22}), we obtain 
\begin{eqnarray*}
   T & = & \varphi\left(\frac{\gamma}{\beta \tilde{S}
           e^{(\beta/\gamma)\tilde{R}}} \right)  \\
     & = & \frac{\beta}{\gamma}\tilde{S}e^{(\beta/\gamma)\tilde{R}}
           \int_{\gamma/(\beta \tilde{S}
           e^{(\beta/\gamma)\tilde{R}})}^{e^{-(\beta/\gamma)\tilde{R}}} 
           \frac{d\xi}{\psi(\xi)} 
           + \frac{1}{\gamma}\left(\log\,(\beta \tilde{I}) 
           - \log\,\bigl(\beta F(N,\tilde{S},\tilde{R},\beta,\gamma)\bigr) 
          \right) 
\end{eqnarray*}
because of  
$$
   \psi\left(\frac{\gamma}{\beta \tilde{S}e^{(\beta/\gamma)\tilde{R}}} \right) 
   = \beta F(N,\tilde{S},\tilde{R},\beta,\gamma). 
$$
It follows from (\ref{ny15}) that 
$$
   \psi'(\xi) = - \beta \tilde{S}e^{(\beta/\gamma)\tilde{R}} 
                + \frac{\gamma}{\xi}, 
$$
and that 
$\psi'\bigl(\gamma/(\beta \tilde{S}e^{(\beta/\gamma)\tilde{R}})\bigr) = 0$, 
$\psi(\xi)$ is decreasing on 
$[\gamma/(\beta \tilde{S}e^{(\beta/\gamma)\tilde{R}}), 
e^{-(\beta/\gamma)\tilde{R}}]$, and 
$\psi(e^{-(\beta/\gamma)\tilde{R}}) = \beta \tilde{I}$. 
Then we get 
$$
   \beta \tilde{I} \leq \psi(\xi) \leq 
   \beta F(N,\tilde{S},\tilde{R},\beta,\gamma), 
$$
and hence
$$
   \frac{1}{\beta F(N,\tilde{S},\tilde{R},\beta,\gamma)} 
   \leq \frac{1}{\psi(\xi)} 
   \leq \frac{1}{\beta \tilde{I}}. 
$$
Integrating the above inequality over 
$[\gamma/(\beta \tilde{S}e^{(\beta/\gamma)\tilde{R}}), 
e^{-(\beta/\gamma)\tilde{R}}]$, and then multiplying by 
$(\beta/\gamma)\tilde{S}e^{(\beta/\gamma)\tilde{R}}$, 
we are led to 
$$
   \frac{(\beta/\gamma)\tilde{S}-1}
         {\beta F(N,\tilde{S},\tilde{R},\beta,\gamma)} 
   \leq \frac{\beta}{\gamma}\tilde{S}e^{(\beta/\gamma)\tilde{R}}
        \int_{\gamma/(\beta \tilde{S}e^{(\beta/\gamma)\tilde{R}})}
        ^{e^{-(\beta/\gamma)\tilde{R}}} \frac{d\xi}{\psi(\xi)} 
   \leq \frac{(\beta/\gamma)\tilde{S}-1}
         {\beta \tilde{I}}, 
$$
which yields the desired inequality. 
\qed

\begin{thm} \label{ny:thm4}
We observe that $R(\infty) = \alpha$, 
\begin{equation}
   R(\infty) = N - \tilde{S}e^{(\beta/\gamma)\tilde{R}}
       e^{-(\beta/\gamma)R(\infty)}  \label{ny26}
\end{equation}
and that $R(t)$ is an increasing function on $[0,\infty)$ such that 
$$
   \tilde{R} \leq R(t) < \alpha = R(\infty). 
$$
\end{thm}

{\Proof}
It follows that 
\begin{eqnarray*}
   R(\infty) = \lim_{t \to \infty} R(t) 
   & = & \lim_{t \to \infty} 
         - \frac{\gamma}{\beta}\log\,\varphi^{-1}(t) \\
   & = & \lim_{u \to e^{-(\beta/\gamma)\alpha}+0} 
         - \frac{\gamma}{\beta}\log\,u  \\
   & = & \alpha.
\end{eqnarray*}
Since $\alpha = R(\infty)$, the identity (\ref{ny26}) 
follows from the definition of $\alpha$. 
Since 
$e^{-(\beta/\gamma)\alpha} < \varphi^{-1}(t) \leq e^{-(\beta/\gamma)\tilde{R}}$, 
we find that 
$$
   - \frac{\gamma}{\beta} \log\,e^{-(\beta/\gamma)\tilde{R}} 
   \leq R(t) < - \frac{\gamma}{\beta} \log\,e^{-(\beta/\gamma)\alpha}, 
$$
or
$$
   \tilde{R} \leq R(t) < \alpha = R(\infty). 
$$
It is clear that $R(t)$ is increasing on $[0,\infty)$ in view of the fact that 
$\varphi^{-1}(t)$ is decreasing $[0,\infty)$. 
\qed

\begin{thm} \label{ny:thm5}
We find that
\begin{equation}
   S(\infty) = \tilde{S}e^{(\beta/\gamma)\tilde{R}}
       e^{-(\beta/\gamma)R(\infty)}  \label{ny27}
\end{equation}
and that $S(t)$ is a decreasing function on $[0,\infty)$ such that 
$$
       \tilde{S} \geq S(t) > \tilde{S}e^{(\beta/\gamma)\tilde{R}}
       e^{-(\beta/\gamma)\alpha} 
       = S(\infty).
$$
\end{thm}

{\Proof}
The identity (\ref{ny27}) follows from 
\begin{eqnarray*}
   S(\infty) = \lim_{t \to \infty} S(t) 
   & = & \lim_{t \to \infty} 
         \tilde{S}e^{(\beta/\gamma)\tilde{R}}\varphi^{-1}(t) \\
   & = & \lim_{u \to e^{-(\beta/\gamma)\alpha}+0} 
         \tilde{S}e^{(\beta/\gamma)\tilde{R}}u \\
   & = & \tilde{S}e^{(\beta/\gamma)\tilde{R}}
         e^{-(\beta/\gamma)\alpha} \\
   & = & \tilde{S}e^{(\beta/\gamma)\tilde{R}}
         e^{-(\beta/\gamma)R(\infty)}. 
\end{eqnarray*}
Since 
$e^{-(\beta/\gamma)\alpha} < \varphi^{-1}(t) \leq e^{-(\beta/\gamma)\tilde{R}}$, we obtain 
$$
   \tilde{S}e^{(\beta/\gamma)\tilde{R}}
       e^{-(\beta/\gamma)\alpha} 
   < \tilde{S}e^{(\beta/\gamma)\tilde{R}}\varphi^{-1}(t) 
   \leq \tilde{S}e^{(\beta/\gamma)\tilde{R}}
       e^{-(\beta/\gamma)\tilde{R}}. 
$$
Hence we get 
$$
   \tilde{S}e^{(\beta/\gamma)\tilde{R}}
       e^{-(\beta/\gamma)\alpha}
   < S(t) \leq \tilde{S}. 
$$
Since $\varphi^{-1}(t)$ is decreasing on $[0,\infty)$, we see that 
$S(t)$ is also decreasing on $[0,\infty)$. 
\qed

\begin{thm} \label{ny:thm6} 
If 
\begin{equation}
   \tilde{S} \leq 
   \frac{\gamma}{\beta} 
   + \frac{1}{2}\left(\tilde{I} + 
     \sqrt{\frac{4\gamma}{\beta}\tilde{I} + \tilde{I}^{2}}\right), 
   \label{ny28}
\end{equation}
then there exists a number $T_{1}$\ $(T < T_{1})$ such that 
$I(t)$ is concave in $(0,T_{1})$, and is convex in $(T_{1},\infty)$. 
If 
\begin{equation}
   \tilde{S} > \frac{\gamma}{\beta} 
               + \frac{1}{2}\left(\tilde{I} 
               + \sqrt{\frac{4\gamma}{\beta}\tilde{I} + \tilde{I}^{2}}\right), 
      \label{ny29}
\end{equation}
then there exist two numbers $T_{2}$ and $T_{3}$\ $(0 < T_{2} < T < T_{3})$ 
such that $I(t)$ is convex in $(0,T_{1}) \cup (T_{3},\infty)$, and is concave 
in $(T_{2},T_{3})$ {\rm ({\it cf}. {\sc Figures} 3 and 4)}. 
\end{thm}

%Figure 3
\begin{figure}[h]
  \begin{center}
    \includegraphics[height=6cm]{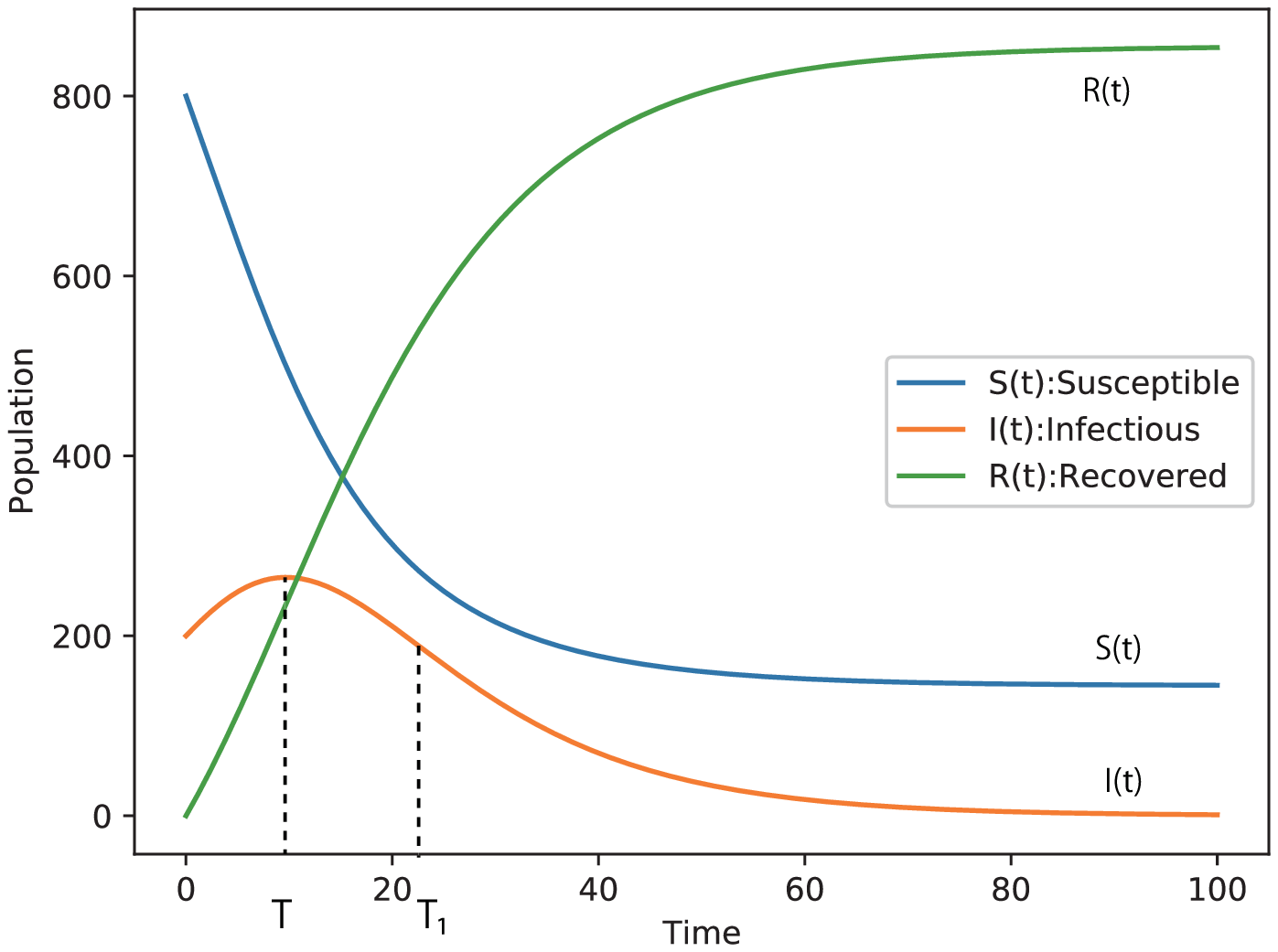}
    \caption{Variation of $S(t)$, $I(t)$ and $R(t)$ 
obtained by the numerical integration of the initial value problem (\ref{ny1})--(\ref{ny4}) for $N=1000, \tilde{S}=800, \tilde{I}=200, \tilde{R}=0, \beta=0.2/1000$ and $\gamma=0.1$. In this case 
$\tilde{S}(=800) > \gamma/\beta(=500) > S(\infty)(=144.56\cdots)$ and 
the condition (\ref{ny28}) is satisfied because $\tilde{S}(=800) < 
\gamma/\beta + (1/2)\left(\tilde{I}+\sqrt{(4\gamma/\beta)\tilde{I}+\tilde{I}^{2}}\right)(= 931.66\cdots)$. }
  \end{center}
\end{figure}

{\Proof}
First note that the hypotheses (A$_{1}$) and (A$_{4}'$) imply that 
$$
   \tilde{S} > \frac{\gamma}{\beta} > N - \alpha = N - R(\infty) = S(\infty)
$$
because $I(\infty) = 0$ and $R(\infty) = \alpha$. 
Differentiating (\ref{ny25}) with respect to $t$ and taking 
(\ref{ny1}), (\ref{ny2}) into account, we have
\begin{eqnarray*}
   I''(t) 
   & = & \bigl(\beta S'(t)\bigr)I(t) + (\beta S(t) - \gamma)I'(t) \\
   & = & \beta (- \beta S(t)I(t))I(t) 
         + (\beta S(t) - \gamma)(\beta S(t)I(t) - \gamma I(t)) \\
   & = & \left(- \beta^{2}S(t)I(t) 
         + \beta^{2}S(t)^{2} - 2 \beta\gamma S(t) + \gamma^{2}\right)I(t) \\
   & = & \beta^{2}\left(S(t)^{2} - S(t)I(t) - \frac{2\gamma}{\beta}S(t) 
         + \frac{\gamma^{2}}{\beta^{2}}\right)I(t) \\
   & = & \beta^{2}\left(S(t) - \frac{2\gamma}{\beta} 
         + \frac{\gamma^{2}}{\beta^{2}}\frac{1}{S(t)}
         - I(t) \right)S(t)I(t). 
\end{eqnarray*}
Now we investigate the sign of $I''(t)$. 
We define 
$$
   G(t) := S(t) - \frac{2\gamma}{\beta} 
         + \frac{\gamma^{2}}{\beta^{2}}\frac{1}{S(t)}
         - I(t) 
$$
and differentiate both sides of the above with respect to $t$ to obtain 
\begin{eqnarray*}
   G'(t) & = & 
   S'(t) - \frac{\gamma^{2}}{\beta^{2}}\frac{S'(t)}{S(t)^{2}} - I'(t) \\
   & = & S'(t) 
         - \frac{\gamma^{2}}{\beta^{2}}\,\frac{-\beta S(t)I(t)}{S(t)^{2}} 
         - I'(t) \\
   & = & - \beta S(t)I(t) + \frac{\gamma^{2}}{\beta}\frac{I(t)}{S(t)} 
         - \bigl(\beta S(t)I(t)-\gamma I(t)\bigr) \\
   & = & - 2 \beta S(t)I(t) +  \frac{\gamma^{2}}{\beta}\frac{I(t)}{S(t)} 
         + \gamma I(t) \\
   & = & -2\beta \left(S(t)^{2} - \frac{\gamma}{2\beta}S(t) 
         - \frac{\gamma^{2}}{2\beta^{2}}\right)\frac{I(t)}{S(t)} \\
   & = & - 2\beta \left(S(t)-\frac{\gamma}{\beta}\right)
         \left(S(t)+\frac{\gamma}{2\beta}\right)\frac{I(t)}{S(t)}. 
\end{eqnarray*}
Since $S(t)+(\gamma/(2\beta)) > 0$, 
we find that $G'(t) = 0$ for 
$t = T = S^{-1}\bigl(\gamma/\beta\bigr)$, and that 
$G'(t) < 0\ [\mbox{resp.}\ > 0]$ if $t < T [\mbox{resp.}\ > T]$. 
Therefore, $G(t)$ is decreasing in $[0, T)$ and increasing in $(T,\infty)$. 
It can be shown that 
\begin{eqnarray*}
   G(0) & = & \tilde{S} - \frac{2\gamma}{\beta} 
              + \frac{\gamma^{2}}{\beta^{2}}\frac{1}{\tilde{S}} - \tilde{I} \\
   & = & \frac{1}{\tilde{S}}\left(\tilde{S}^{2} 
         - \left(\frac{2\gamma}{\beta} + \tilde{I}\right)\tilde{S} 
         + \frac{\gamma^{2}}{\beta^{2}}\right) \\
   & = & \frac{1}{\tilde{S}}(\tilde{S} - s_{1})(\tilde{S} - s_{2}), 
\end{eqnarray*}
where
\begin{eqnarray*}
   s_{1} & = & \frac{\gamma}{\beta} 
               + \frac{1}{2}\left(\tilde{I} 
               - \sqrt{\frac{4\gamma}{\beta}\tilde{I} + \tilde{I}^{2}}\right)
               \ \left( < \frac{\gamma}{\beta}\right), \\   
   s_{2} & = & \frac{\gamma}{\beta} 
               + \frac{1}{2}\left(\tilde{I} 
               + \sqrt{\frac{4\gamma}{\beta}\tilde{I} + \tilde{I}^{2}}\right)
               \ \left( > \frac{\gamma}{\beta}\right). 
\end{eqnarray*}
Moreover we see that
\begin{eqnarray*}
   G(T) & = & S(T) - \frac{2\gamma}{\beta} 
              + \frac{\gamma^{2}}{\beta^{2}}\frac{1}{S(T)} - I(T) \\
   & = & \frac{\gamma}{\beta} - \frac{2\gamma}{\beta} 
         + \frac{\gamma^{2}}{\beta^{2}}\frac{\beta}{\gamma} 
         - \max_{t \geq 0} I(t) \\
   & = & - \max_{t \geq 0} I(t) < 0, 
\end{eqnarray*}
and that
\begin{eqnarray*}
   \lim_{t \to \infty} G(t) 
   & = & S(\infty) - \frac{2\gamma}{\beta} 
         + \frac{\gamma^{2}}{\beta^{2}}\frac{1}{S(\infty)} - I(\infty) \\
   & = & \frac{1}{S(\infty)}\left(\frac{\gamma}{\beta} - S(\infty)\right)^{2} 
         > 0. 
\end{eqnarray*}
If (\ref{ny28}) is satisfied, then $G(0) \leq 0$, and hence there exists 
a number $T_{1} > T$ such that $G(T_{1}) = 0$, $G(t)$ is negative in 
$(0,T_{1})$, and $G(t)$ is positive in $(T_{1},\infty)$. 
Since $I''(t) = \beta^{2}G(t)S(t)I(t)$, we conclude that 
$I(t)$ is concave in $(0,T_{1})$, and is convex in $(T_{1},\infty)$.
If (\ref{ny29}) is satisfied, then $G(0) > 0$, and therefore there exist 
two numbers $T_{2}$ and $T_{3}$\ $(0 < T_{2} < T < T_{3})$ 
such that $G(T_{2}) = G(T_{3}) = 0$, 
$G(t)$ is positive in $(0,T_{2}) \cup (T_{3},\infty)$, 
and $G(t)$ is negative in $(T_{2},T_{3})$. Consequently we see that 
$I(t)$ is convex in $(0,T_{2}) \cup (T_{3},\infty)$, and is concave 
in $(T_{2},T_{3})$.
\qed

%Figure 4
\begin{figure}[h]
  \begin{center}
    \includegraphics[height=6cm]{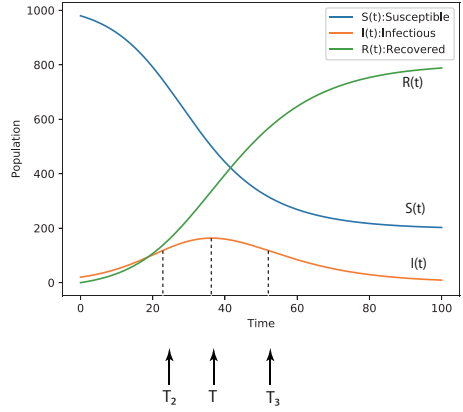}
    \caption{Variation of $S(t)$, $I(t)$ and $R(t)$ 
obtained by the numerical integration of the initial value problem (\ref{ny1})--(\ref{ny4}) for $N=1000, \tilde{S}=980, \tilde{I}=20, \tilde{R}=0, 
\beta=0.2/1000$ and $\gamma=0.1$. In this case 
$\tilde{S}(=980) > \gamma/\beta(=500) > S(\infty)(=196.46\cdots)$ and 
the condition (\ref{ny29}) is satisfied because 
$\tilde{S}(=980) > \gamma/\beta + (1/2)\left(\tilde{I}+\sqrt{(4\gamma/\beta)\tilde{I}+\tilde{I}^{2}}\right)(= 581.41\cdots)$. }
  \end{center}
\end{figure}

\begin{thm} \label{ny:thm7}
We get the following identity: 
$$
  S(\infty)  
 = \tilde{S} + \tilde{I} + \frac{\gamma}{\beta}\log \frac{S(\infty)}{\tilde{S}}.
$$
\end{thm}

{\Proof}
Since $I(\infty) = 0$, it is easily verified that 
\begin{eqnarray*}
   S(\infty)
   & = & N - \tilde{R} + \tilde{R} - R(\infty) \\
   & = & \tilde{S} + \tilde{I} 
         + \frac{\gamma}{\beta}\left(\frac{\beta}{\gamma}\tilde{R}
         - \frac{\beta}{\gamma}R(\infty)\right) \\
   & = & \tilde{S} + \tilde{I}
         + \frac{\gamma}{\beta}\,\log 
         \left(e^{(\beta/\gamma)\tilde{R}}
         e^{-(\beta/\gamma)R(\infty)}\right) \\
   & = & \tilde{S} + \tilde{I} 
         + \frac{\gamma}{\beta}\,\log \frac{S(\infty)}{\tilde{S}} 
\end{eqnarray*}
in light of (\ref{ny27}). 
\qed

\begin{thm} \label{ny:thm8}
We observe that
$$
   S'(\infty) = I'(\infty) = R'(\infty) = 0. 
$$
\end{thm}

{\Proof}
Since $I(\infty) = 0$, we conclude from (\ref{ny1})--(\ref{ny3}) that 
\begin{eqnarray*}
   S'(\infty) & = & - \beta S(\infty)I(\infty) = 0, \\
   I'(\infty) & = & \beta S(\infty)I(\infty) - \gamma I(\infty)  = 0, \\
   R'(\infty) & = & \gamma I(\infty) = 0. 
\end{eqnarray*}
\qed

\begin{thm} \label{ny:thm9} 
Let $(S(t), I(t), R(t))$ be the exact solution 
{\rm (\ref{ny18})--(\ref{ny20})} of the initial value problem 
{\rm (\ref{ny1})--(\ref{ny4})}, and let 
$$
   \bigl(\hat{S}(u),\hat{I}(u),\hat{R}(u)\bigr) 
   := \bigl(S(\varphi(u)),I(\varphi(u)),R(\varphi(u))\bigr). 
$$
Then $\bigl(\hat{S}(u),\hat{I}(u),\hat{R}(u)\bigr)$ is a solution of 
the initial value problem for the linear differential system 
\begin{eqnarray}
   \frac{d\hat{S}(u)}{du} & = & \frac{\hat{S}(u)}{u}, \label{ny30} \\
   \frac{d\hat{I}(u)}{du} & = & - \frac{\hat{S}(u)}{u} 
                                + \frac{\gamma}{\beta}\,\frac{1}{u}, 
                                \label{ny31}\\
   \frac{d\hat{R}(u)}{du} & = & - \frac{\gamma}{\beta}\,\frac{1}{u} 
                                \label{ny32}
\end{eqnarray}
for $u \in (e^{-(\beta/\gamma)\alpha}, e^{-(\beta/\gamma)\tilde{R}})$, 
subject to the initial condition
\begin{eqnarray}
   \hat{S}\left(e^{-(\beta/\gamma)\tilde{R}}\right) & = & \tilde{S}, 
                                \label{ny33}  \\
   \hat{I}\left(e^{-(\beta/\gamma)\tilde{R}}\right) & = & \tilde{I}, 
                                \label{ny34}  \\
   \hat{R}\left(e^{-(\beta/\gamma)\tilde{R}}\right) & = & \tilde{R}. 
                                \label{ny35} 
\end{eqnarray}
\end{thm}

{\Proof}
First we note that 
\begin{equation}
   \hat{I}(u) = I(\varphi(u)) = \frac{1}{\beta}\psi(u) \label{ny36}
\end{equation}
in light of (\ref{ny24}). 
It follows from (\ref{ny1}) that 
$$
   S'(\varphi(u)) = - \beta S(\varphi(u))I(\varphi(u)) 
   = - \beta \hat{S}(u)\hat{I}(u). 
$$
Then we arrive at
\begin{eqnarray*}
   \frac{d\hat{S}(u)}{du} & = & 
         \frac{dS(t)}{dt}\Big\vert_{t=\varphi(u)}\varphi'(u)
         =  S'(\varphi(u))\left(- \frac{1}{u\psi(u)}\right) \\
   & = & \left(- \beta \hat{S}(u)\hat{I}(u)\right)
         \left(- \frac{1}{u\psi(u)}\right) 
         =  \beta \hat{S}(u)\hat{I}(u)\left(\frac{1}{u\psi(u)}\right) \\
   & = & \frac{\hat{S}(u)}{u}
\end{eqnarray*}
in view of (\ref{ny36}). Analogously we have
\begin{eqnarray*}
   \frac{d\hat{R}(u)}{du} & = & 
         \frac{dR(t)}{dt}\Big\vert_{t=\varphi(u)}\varphi'(u)
         =  R'(\varphi(u))\left(- \frac{1}{u\psi(u)}\right) \\
   & = & \left( \gamma \hat{I}(u)\right)
         \left(- \frac{1}{u\psi(u)}\right) \\
   & = & - \frac{\gamma}{\beta} \frac{1}{u}
\end{eqnarray*}
and 
\begin{eqnarray*}
   \frac{d\hat{I}(u)}{du} & = & 
         \frac{dI(t)}{dt}\Big\vert_{t=\varphi(u)}\varphi'(u)
         =  I'(\varphi(u))\left(- \frac{1}{u\psi(u)}\right) \\
   & = & \left(\beta \hat{S}(u)\hat{I}(u) - \gamma \hat{I}(u)\right)
         \left(- \frac{1}{u\psi(u)}\right) \\
   & = & - \frac{\hat{S}(u)}{u} + \frac{\gamma}{\beta} \frac{1}{u}.
\end{eqnarray*}
It is clear that
$$
   \hat{S}\left(e^{-(\beta/\gamma)\tilde{R}}\right) 
   =  S\left(\varphi\left(e^{-(\beta/\gamma)\tilde{R}}\right)\right) \\  
   =  S(0) = \tilde{S}, 
$$
$$
   \hat{I}\left(e^{-(\beta/\gamma)\tilde{R}}\right) 
   =  I\left(\varphi\left(e^{-(\beta/\gamma)\tilde{R}}\right)\right) \\  
   =  I(0) = \tilde{I}, 
$$
$$
   \hat{R}\left(e^{-(\beta/\gamma)\tilde{R}}\right) 
   =  R\left(\varphi\left(e^{-(\beta/\gamma)\tilde{R}}\right)\right) \\  
   =  R(0) = \tilde{R}. 
$$
Therefore, $\bigl(\hat{S}(u),\hat{I}(u),\hat{R}(u)\bigr)$ 
is a solution of the initial value problem (\ref{ny30})--(\ref{ny35}). 
\qed

\begin{thm} \label{ny:thm10}
Solving the initial value problem {\rm (\ref{ny30})--(\ref{ny35})}, 
we are led to the solution {\rm (\ref{ny8})--(\ref{ny10})} 
for $u \in (e^{-(\beta/\gamma)\alpha}, e^{-(\beta/\gamma)\tilde{R}}]$. 
\end{thm}

{\Proof}
Since (\ref{ny30}) is equivalent to 
$$
   \frac{d}{du}\left(\frac{1}{u}\hat{S}(u)\right) = 0, 
$$
we have 
$$
   \hat{S}(u) = ku 
$$
for some constant $k$.  It follows from (\ref{ny33}) that
$$
   \hat{S}\left(e^{-(\beta/\gamma)\tilde{R}}\right) 
   = ke^{-(\beta/\gamma)\tilde{R}} = \tilde{S}, 
$$
and therefore
$$
   k = \tilde{S}e^{(\beta/\gamma)\tilde{R}},
$$
which implies 
$$
   \hat{S}(u) = \tilde{S}e^{(\beta/\gamma)\tilde{R}}u. 
$$
Solving (\ref{ny32}) yields 
$$
   \hat{R}(u) = - \frac{\gamma}{\beta}\log u + k
$$
for some constant $k$. The initial condition (\ref{ny35}) implies 
$$
   \hat{R}\left(e^{-(\beta/\gamma)\tilde{R}}\right) 
   = - \frac{\gamma}{\beta}\log e^{-(\beta/\gamma)\tilde{R}} + k 
   = \tilde{R} + k = \tilde{R}, 
$$
and hence $k = 0$. Consequently we obtain 
$$
   \hat{R}(u) = - \frac{\gamma}{\beta}\log u.
$$
Since 
$$
   \frac{\hat{S}(u)}{u} = \tilde{S}e^{(\beta/\gamma)\tilde{R}},
$$
we have
$$
   \frac{d\hat{I}(u)}{du}  
   =  - \tilde{S}e^{(\beta/\gamma)\tilde{R}} 
      + \frac{\gamma}{\beta}\,\frac{1}{u}.
$$
Hence we derive 
$$
   \hat{I}(u) = - \tilde{S}e^{(\beta/\gamma)\tilde{R}}u 
   + \frac{\gamma}{\beta} \log u + k 
$$
for some constant $k$. From the initial condition (\ref{ny34}) we see that 
$$
   \hat{I}\left(e^{-(\beta/\gamma)\tilde{R}}\right) 
   = - \tilde{S} + \frac{\gamma}{\beta} \log e^{-(\beta/\gamma)\tilde{R}} + k 
   = - \tilde{S} - \tilde{R} + k = \tilde{I},
$$
which implies  
$$
   k = \tilde{I} + \tilde{S} +  \tilde{R} = N.
$$
Therefore we have 
$$
   \hat{I}(u) = - \tilde{S}e^{(\beta/\gamma)\tilde{R}}u 
   + \frac{\gamma}{\beta} \log u + N. 
$$
\qed

%Figure 5
\begin{figure}[ht]
  \begin{center}
    \includegraphics[height=5.5cm]{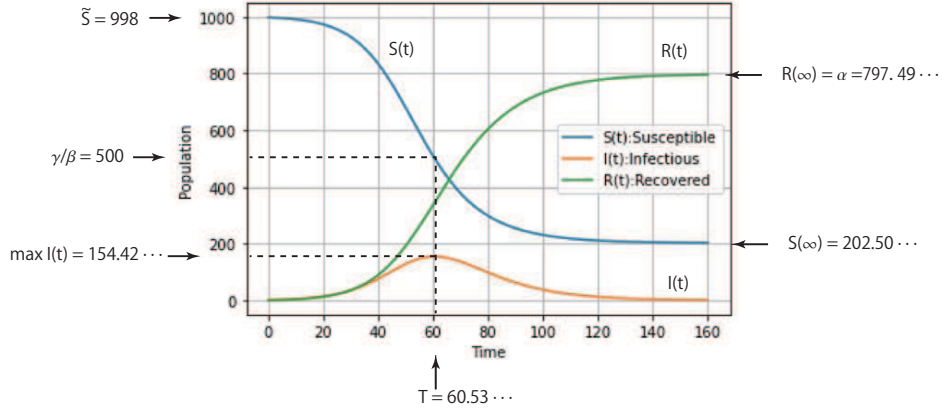}
    \caption{Variation of $S(t), I(t), R(t)$ obtained by the numerical 
    integration of the initial value problem (\ref{ny1})--(\ref{ny4}) for 
    $N=1000, \tilde{S}=998, \tilde{I}=2, \tilde{R}=0, \beta=0.2/1000$ and 
    $\gamma=0.1$. 
    In this case we obtain $R(\infty)=\alpha=797.49\cdots$, 
    $I(\infty)=0$, $S(\infty)=N-R(\infty)=202.50\cdots$, 
    $\gamma/\beta =500$, $\tilde{S}(=998) > \gamma/\beta(=500) 
    > S(\infty)(=202.50\cdots)$, $T=60.53\cdots$ and 
    $\max_{t\geq 0} I(t) = 154.42\cdots$, where 
    $T$ is calculated by }
\begin{eqnarray*}
   T & = & \varphi\bigl(\gamma/(\beta\tilde{S})\bigr) 
           =  \int_{500/998}^{1} \frac{d\xi}{\xi\psi(\xi)} \\
     & = & \int_{500/998}^{1} \frac{d\xi}{\xi(0.2-(0.2/1000)\times 998\,\xi
           +0.1\log\xi)} 
           =  60.53\cdots.
\end{eqnarray*}
  \end{center}
\end{figure}

\begin{rem} \label{ny:rem1} \rm
The hypothesis (A$_{3}$) is satisfied if $\tilde{R} = 0$, 
since $N > \tilde{S}$. 
\end{rem}

\begin{rem} \label{ny:rem2} \rm
When $\tilde{R} = 0$, the first order 
differential equation (\ref{ny6}) was studied by 
Kermack and McKendrick \cite[p.713]{km27}. 
\end{rem}

\begin{rem} \label{ny:rem3} \rm
We differentiate the first order differential 
equation (\ref{ny6}) 
to obtain the following second order differential equation 
\begin{eqnarray*}
   R''(t) & = & \gamma\left(- \tilde{S}e^{(\beta/\gamma)\tilde{R}}
                \left(-\frac{\beta}{\gamma}\right)R'(t)
                e^{-(\beta/\gamma)R(t)} - R'(t)\right) \\
   & = & \tilde{S}e^{(\beta/\gamma)\tilde{R}}\beta R'(t)
         e^{-(\beta/\gamma)R(t)} - \gamma R'(t), 
\end{eqnarray*}
which was investigated by Harko, Lobo and Mak \cite{hlm14}. 
\end{rem}

\begin{rem} \label{ny:rem4} \rm 
The right differential coefficient $I_{+}'(0)$ is positive because 
\begin{eqnarray*}
   I_{+}'(0) 
   & = & \lim_{t \to +0} I'(t) 
         = \lim_{t \to +0} \bigl(\beta S(t)I(t) - \gamma I(t)\bigr) \\
   & = & \beta \tilde{S}\tilde{I} - \gamma \tilde{I} 
         = \bigl(\beta \tilde{S} - \gamma\bigr)\tilde{I} > 0 
\end{eqnarray*}
in view of the hypotheses (A$_{1}$) and (A$_{2}$). 
\end{rem}

\begin{rem} \label{ny:rem5} \rm 
In the case where $(S(\infty) <)\,\tilde{S} \leq \gamma/\beta$ (i.e., 
$I_{+}'(0) \leq 0$) we deduce that $\psi(\xi)$ is increasing in 
$\bigl(e^{-(\beta/\gamma)\alpha}, e^{-(\beta/\gamma)\tilde{R}}\bigr]$, 
$\lim_{\xi \to e^{-(\beta/\gamma)\alpha}+0} \psi(\xi) = 0$ and 
$\psi\bigl(e^{-(\beta/\gamma)\tilde{R}}\bigr) = \beta \tilde{I}$. 
Since $\varphi^{-1}(t)$ is decreasing on $[0,\infty)$, 
$\varphi^{-1}(0) = e^{-(\beta/\gamma)\tilde{R}}$ and 
$\lim_{t \to \infty} \varphi^{-1}(t) = e^{-(\beta/\gamma)\alpha}$, 
it follows that $I(t) = (1/\beta)\psi\bigl(\varphi^{-1}(t)\bigr)$ 
is decreasing on $[0,\infty)$, and that 
$I(0) = (1/\beta)\psi(\varphi^{-1}(0)\bigr) = \tilde{I}$ 
and $I(\infty) = \lim_{t \to \infty} I(t) 
= \lim_{\xi \to e^{-(\beta/\gamma)\alpha}+0} (1/\beta)\psi(\xi) = 0$ 
{\rm ({\it cf}. {\sc Figure} 6)}. 
\end{rem}

%Figure 6
\begin{figure}[ht]
  \begin{center}
    \includegraphics[height=4.5cm]{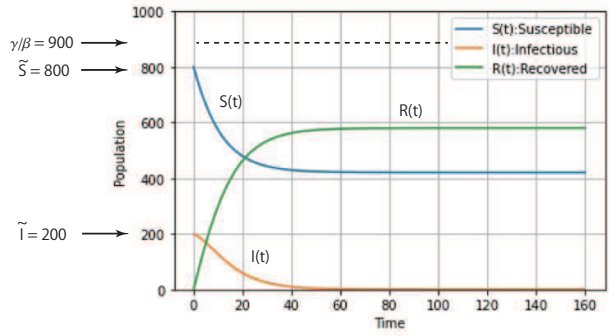}
    \caption{Variation of $S(t), I(t), R(t)$ obtained by the numerical 
    integration of the initial value problem (\ref{ny1})--(\ref{ny4}) for 
    $N=1000, \tilde{S}=800, \tilde{I}=200, \tilde{R}=0, \beta=0.2/1000$ and 
    $\gamma=0.18$. 
    In this case we find that 
    $\gamma/\beta (=900) > \tilde{S} (= 800)$, $I(t)$ is decreasing on 
    $[0,\infty)$ and $I(\infty) = 0$. }
  \end{center}
\end{figure}

\begin{rem} \label{ny:rem6} \rm
The constant $F(N,\tilde{S},\tilde{R},\beta,\gamma)$ defined in 
Corollary \ref{ny:cor1} is equal to $\max\limits_{t \geq 0} I(t)$ given 
in Theorem \ref{ny:thm3}. 
\end{rem}

\begin{rem} \label{ny:rem7} \rm 
It follows from Theorems \ref{ny:thm3}--\ref{ny:thm5} that 
$S(t) > 0$, $I(t) > 0$ for $t \geq 0$ and $R(t) > 0$ for $t > 0$. 
\end{rem}

\begin{rem} \label{ny:rem8} \rm 
If there exists a solution $R(t)$ of the initial value problem for (\ref{ny6}) 
with the initial condition $R(0) = \tilde{R}$, then 
$(S(t),I(t),R(t))$ is a solution of the initial value problem 
{\rm (\ref{ny1})--(\ref{ny4})}, where $S(t)$ and $I(t)$ are given by
\begin{eqnarray*}
   S(t) & = & \tilde{S}e^{(\beta/\gamma)\tilde{R}}e^{-(\beta/\gamma)R(t)}, \\
   I(t) & = & - R(t) - \tilde{S}e^{(\beta/\gamma)\tilde{R}}
              e^{-(\beta/\gamma)R(t)} + N 
\end{eqnarray*}
(cf. Corollary \ref{ny:cor1}). In fact we obtain 
\begin{eqnarray*}
   S'(t) & = & \tilde{S}e^{(\beta/\gamma)\tilde{R}}e^{-(\beta/\gamma)R(t)}
               \left(-\frac{\beta}{\gamma}R'(t)\right) \\
         & = & - \beta \tilde{S}e^{(\beta/\gamma)\tilde{R}}
               e^{-(\beta/\gamma)R(t)}\frac{R'(t)}{\gamma} \\
         & = & - \beta S(t)\bigl(N - \tilde{S}e^{(\beta/\gamma)\tilde{R}}
               e^{-(\beta/\gamma)R(t)} - R(t)\bigr) \\
         & = & - \beta S(t)I(t), \\
   R'(t) & = & \gamma I(t), \\
   I'(t) & = & - R'(t) + \beta S(t)I(t) \\
         & = & \beta S(t)I(t) - \gamma I(t) 
\end{eqnarray*}
and $S(0) = \tilde{S}e^{(\beta/\gamma)\tilde{R}}e^{-(\beta/\gamma)\tilde{R}} = 
\tilde{S}$, $I(0) = -\tilde{R}-\tilde{S} + N = \tilde{I}$. 
\end{rem}

\begin{rem} \label{ny:rem9} \rm 
Let $R(t)$ be given by (\ref{ny20}).  
Then $R(t)$ is a positive and increasing solution of the initial value 
problem for (\ref{ny6}) with the initial condition $R(0) = \tilde{R}$. 
In fact, Theorem \ref{ny:thm4} implies that $R(t)$ is positive and 
increasing in $(0,\infty)$. We observe that 
\begin{eqnarray*}
   R'(t) & = & - \frac{\gamma}{\beta} 
               \frac{\bigl(\varphi^{-1}(t)\bigr)'}{\varphi^{-1}(t)} 
               = - \frac{\gamma}{\beta}\bigl(-\psi(\varphi^{-1}(t))\bigr) 
               = \frac{\gamma}{\beta}\psi(\varphi^{-1}(t)) \\
         & = & \frac{\gamma}{\beta}
               \left(\beta N - \beta \tilde{S}e^{(\beta/\gamma)\tilde{R}}
               \varphi^{-1}(t) + \gamma \log \varphi^{-1}(t) \right) \\
         & = & \gamma \left(
               N - \tilde{S}e^{(\beta/\gamma)\tilde{R}}
               \varphi^{-1}(t) + \frac{\gamma}{\beta} \log \varphi^{-1}(t)
               \right) \\
         & = & \gamma \left(
               N - \tilde{S}e^{(\beta/\gamma)\tilde{R}}e^{-(\beta/\gamma)R(t)} 
               - R(t) \right) \ (t > 0) 
\end{eqnarray*}
in view of (\ref{ny15}) and (\ref{ny23}). 
It is easy to check that $R(0) = - (\gamma/\beta)\log \varphi^{-1}(0) 
= - (\gamma/\beta) \log e^{-(\beta/\gamma)\tilde{R}} = \tilde{R}$. 
Therefore, $R(t)$ satisfies (\ref{ny6}) and the initial condition 
$R(0) = \tilde{R}$. 
\end{rem}

\begin{rem} \label{ny:rem10} \rm 
Let $R(t)$ be given by (\ref{ny20}).  
Then,  Remark \ref{ny:rem9} implies that 
$R(t)$ is a solution of the initial value problem for (\ref{ny6}) 
with the initial condition $R(0) = \tilde{R}$. 
It follows from Remark \ref{ny:rem8} that 
\begin{eqnarray*}
   S(t) & = & \tilde{S}e^{(\beta/\gamma)\tilde{R}}
              e^{-(\beta/\gamma)(-(\gamma/\beta)\log \varphi^{-1}(t))} \\
        & = & \tilde{S}e^{(\beta/\gamma)\tilde{R}}\varphi^{-1}(t), \\
   I(t) & = & \frac{\gamma}{\beta}\log \varphi^{-1}(t) 
              - \tilde{S}e^{(\beta/\gamma)\tilde{R}}\varphi^{-1}(t) + N, \\
   R(t) & = & -\frac{\gamma}{\beta} \log \varphi^{-1}(t)
\end{eqnarray*}
is a solution of the initial value problem {\rm (\ref{ny1})--(\ref{ny4})}, 
which is equal to the solution {\rm (\ref{ny18})--(\ref{ny20})} 
in Theorem \ref{ny:thm2}. 
\end{rem}

%\subsection*{Acknowledgements}
%The author would like to thank the referee for her/his helpful comments and 
%suggestions. 

\end{document}